# "Prediction of enthalpy for nitrogen gas" revised

I.H. Umirzakov

Institute of Thermophysics, Lavrentev prospect, 1, 630090 Novosibirsk, Russia

e-mail: cluster125@gmail.com

**Abstract**

It is shown that the energy spectrum of the pure vibrational levels of the molecule consisting of two atoms interacting with each other via the modified Rosen-Morse potential, the analytical expressions for the vibrational partition function and enthalpy of the diatomic molecule, obtained in the paper "Prediction of enthalpy for nitrogen gas, M. Deng, C.S. Jia, Evr. Phys. J. Plus **133,** 258 (2018)", are incorrect.

**Introduction**

The modified Rosen-Morse potential is given by [1]

$$V(r) = D_e \left(1 - \frac{e^{\alpha r_e} + q}{e^{\alpha r} + q}\right)^2, \qquad (1)$$

where $D_e$ is the dissociation energy, $D_e > 0$, $r$ is the interatomic separation, $r \geq 0$, $r_e$ is the equilibrium bond length, $r_e > 0$, $q$ and $\alpha$ are the adjustable parameters, $\alpha > 0$ and $q > 0$. Here the energy is measured from the minimum of the potential well, so $V(r_e) = 0$ and $V(r \to \infty) = D_e$.

The expression for the energy spectrum of the pure vibrational levels of the molecule consisting of two atoms interacting with each other via the modified Rosen-Morse potential was used in [1] to obtain analytical expressions for the vibrational partition function. The expression for the enthalpy of the ideal gas of the diatomic molecules was obtained using the vibrational partition function and rigid rotator model in order to take into account the rotations of the molecule. The calculated enthalpy for the nitrogen molecule $N_2$ was compared with the experimental data.

We show in the present paper that the energy spectrum of the pure vibrational levels of the diatomic molecule and the analytical expressions for the vibrational partition function and enthalpy of the diatomic molecule obtained in [1] are incorrect.

**Main consideration**

1. According to [1] the energy $E(v)$ of the pure vibrational level with the vibrational quantum number $v$ is defined from

$$E(v) = D_e - \frac{\hbar^2 \alpha^2}{2\mu}(A - B)^2, \qquad (2)$$

$$A = \frac{2\mu D_e}{\hbar^2 \alpha^2 q^2} \cdot \frac{e^{2\alpha r_e} - q^2}{2v + 1 - \sqrt{1 + \frac{8\mu D_e}{\hbar^2 \alpha^2 q^2}\left(e^{\alpha r_e} + q\right)^2}}, \tag{3}$$

$$B = \frac{1}{4}\left(2v + 1 - \sqrt{1 + \frac{8\mu D_e}{\hbar^2 \alpha^2 q^2}\left(e^{\alpha r_e} + q\right)^2}\right), \tag{4}$$

where $\hbar$ is the Planck's constant, $\mu$ is the reduced mass and $q > 0$.

According to [1] Eq. 2 is valid for the vibrational quantum number $v$ which obeys the condition

$$0 \leq v \leq v_{\max}, \tag{5}$$

$$v_{\max} = \left[\left(\sqrt{1 + \gamma(a+1)^2} - 1 - \sqrt{\gamma(a^2-1)}\right)/2\right]_{\text{integer}}, \tag{6}$$

where $v_{\max}$ is the most vibration quantum number, $\gamma = 8\mu D_e / \hbar^2 \alpha^2$, $\gamma > 0$ and $a = e^{\alpha r_e}/q$. Using the condition $q > 0$ we obtain from $a = e^{\alpha r_e}/q$ that $a > 0$. Here $[X]_{\text{integer}}$ is equal to $X$ for an integer $X$, and $[X]_{\text{integer}}$ is equal to the greatest integer less than $X$ for a non-integer $X$.

According to [1] the value of $v_{\max}$ given by Eq. 6 can be obtained using Eq. 2 from the condition

$$dE(v)/dv\big|_{v=v_{\max}} = 0. \tag{7}$$

The partition function of the vibrational states ($Q^v$) of the diatomic molecule is equal to

$$Q^v = \sum_{v=0}^{v_{\max}} e^{-E(v)/kT}, \tag{8}$$

where $k$ the is Boltzmann's constant and $T$ is the temperature [1].

Eq. 2 gives $E(v_{\max}) = 0$ in the case when

$$\sqrt{1 + \gamma(a+1)^2} - 1 - \sqrt{\gamma(a^2-1)} = 2n, \tag{9}$$

where $n = 0, 1, 2...$. The value $E(v_{\max}) = 0$ corresponds to the un-bonded state of the diatomic molecule. Therefore Eq. 6 cannot be used in order to calculate the partition function of bonded vibrational states of the diatomic molecule, and it is necessary replace Eq. 6 by the correct one. For example, Eq. 6 can be replaced by

$$v_{\max} = \left[\left(\sqrt{1 + \gamma(a+1)^2} - 1 - \sqrt{\gamma(a^2-1)}\right)/2 - 1/2\right]_{\text{integer}}. \tag{10}$$

in order to avoid the inclusion of the un-bonded state to the partition function.

2. It is easy to see that Eq. 6 is incorrect in the case: a) $0 < a < 1$.

We have from Eq. 6 that $v_{\max} < 0$ in the case: b) $a > (1+\gamma)/(1-\gamma)$ and $\gamma < 1$.

Therefore in the cases a) and b) the solution of the Schrodinger equation, energy spectrum, given by Eq. 2, analytical expressions for the partition function and enthalpy of the diatomic molecule, obtained in [1] from Eq. 2, loss their physical sense.

3. The vibrational partition function was calculated in [1] using the approximate relation

$$Q_{appr}^{v} = \frac{1}{2}[e^{-E(0)/kT} - e^{-E(v_{max}+1)/kT}] + \int_{0}^{v_{max}+1} e^{-E(x)/kT} dx. \tag{11}$$

The enthalpy of the bonded vibrational states is equal to the mean energy of the states [1], so we have $H^{v} = E^{v}$ and $H_{appr}^{v} = E_{appr}^{v}$ for the exact ($H^{v}$) and approximate ($H_{appr}^{v}$) enthalpies of the bonded vibrational states of diatomic molecule.

Using the exact relation $E = kT^2 \partial \ln Q / T$ we obtain from Eqs. 8 and 11

$$E^{v} = H^{v} = \sum_{v=0}^{v_{max}} E(v) e^{-E(v)/kT}, \tag{12}$$

$$E_{appr}^{v} = H_{appr}^{v} = \frac{1}{2}[E(0)e^{-E(0)/kT} - E(v_{max}+1)e^{-E(v_{max}+1)/kT}] + \int_{0}^{v_{max}+1} E(x)e^{-E(x)/kT} dx, \tag{13}$$

for the exact $E$ and approximate $E_{appr}$ mean energies of the bonded vibrational states, respectively.

The enthalpy $H$ of the molecule is equal to the sum of the rotational $H^{rot}$, translational $H^{tr}$ and vibrational $H^{v}$ enthalpies [1]:

$$H = H^{v} + H^{r} + H^{tr}, \tag{14}$$

$$H^{tr} = 5N_{A}kT/2, \tag{15}$$

$$H^{rot} = N_{A}kT \frac{1 - (T/T_{rot})^2/15 - 8(T/T_{rot})^3/315}{1 + (T/T_{rot})/3 + (T/T_{rot})^2/15 + 4(T/T_{rot})^3/315}, \tag{16}$$

where $N_{A}$ is the Avogadro's number and $T_{rot} = \hbar^2 / 2\mu r_{e}^2 k$.

The detailed analysis of the derivation of Eqs. 8 and 15 [1] from Eq. 7 [1] shows that Eqs. 8 and 15 [1] for the partition function $Q_{appr}^{v}$ and enthalpy $H_{appr}^{v}$ are incorrect. For example:

- it is necessary to replace the incorrect $2\delta_1 + \eta_1$ and $2\delta_1 + \eta_2$ by the correct $2\delta_1 + \eta_1^2$ and $2\delta_1 + \eta_2^2$, respectively, in the right hand side of Eqs. 8 and 15 [1];
- the incorrect term $2\delta_1 e^{-2\lambda\delta_1/kT} \cdot erfi(\sqrt{\lambda(2\delta_1 + \eta_2)/kT})$ must be replaced by the correct one $2\delta_1 e^{-2\lambda\delta_1/kT} erfi(\sqrt{\lambda(2\delta_1 + \eta_2^2)/kT})$ in the numerator at the right hand side of Eq. 15 [1];
- the dimensionalities of the arguments of the exponents $e^{-\eta_1/kT}$ and $e^{-\eta_2/kT}$ in Eq. 15 [1] are equal to that of an inverse of an energy while the dimensionalities of the arguments of the exponents must be dimensionless, and etc.

Therefore the data for the enthalpy presented on Fig. 1 [1] are incorrect if they were obtained using Eq. 15 [1].

The correct Eqs. 8 and 15 [1] are:

$$Q_{appr}^{v} = e^{-D_{e}/kT} G/2, \tag{17}$$

$$H_{appr}^{v} = D_{e} + (\lambda \eta_2^2 e^{\lambda \eta_2^2/kT} - \lambda \eta_1^2 e^{\lambda \eta_1^2/kT})/G + kT/2 \cdot [1 - (e^{\lambda \eta_1^2/kT} - e^{\lambda \eta_2^2/kT})/G] +$$
$$+ 2\lambda \delta_1 G^{-1} e^{-2\lambda \delta_1/kT} \sqrt{\pi kT/\lambda} \left[ erfi\left(\sqrt{\lambda(2\delta_1 + \eta_1^2)/kT}\right) - erfi\left(\sqrt{\lambda(2\delta_1 + \eta_2^2)/kT}\right) \right] + \tag{18}$$
$$kTG^{-1}[(\eta_2 + \sqrt{2\delta_1 + \eta_2^2})e^{\lambda \eta_2^2/kT} - (\eta_1 + \sqrt{2\delta_1 + \eta_1^2})e^{\lambda \eta_1^2/kT}],$$

where

$$G = e^{\lambda \eta_1^2 / kT} - e^{\lambda \eta_2^2 / kT} + \sqrt{\pi kT/\lambda}\left[ erfi\left(\sqrt{\lambda \eta_1^2 / kT}\right) - erfi\left(\sqrt{\lambda \eta_2^2 / kT}\right)\right] +$$
$$+ e^{-2\lambda \delta_1 / kT} \sqrt{\pi kT/\lambda}\left[ erfi\left(\sqrt{\lambda (2\delta_1 + \eta_1^2)/kT}\right) - erfi\left(\sqrt{\lambda (2\delta_1 + \eta_2^2)/kT}\right)\right] \quad (19)$$

4. According to [1] $E(0) > 0$ and $dE(v)/dv > 0$, where $0 \leq v \leq v_{max}$, for the nitrogen molecule $N_2$. Therefore we obtain from Eqs. 12-13

$$H^v > 0, \qquad (20)$$

$$H^v_{appr} > 0 \qquad (21)$$

for $T > 0$, and

$$H^v = H^v_{appr} = E(0) > 0 \qquad (22)$$

in the limit $T \to 0$ (see also Fig. 3).

According to Eqs. 15-16 the rotational and translational enthalpies are positive at $T > 0$, and these enthalpies vanish in the limit $T \to 0$ [1]. Therefore we have from Eqs. 12-14 using Eqs. 15-16 and 20-22

$$H > 0, \qquad (23)$$

$$H_{appr} > 0 \qquad (24)$$

for $T > 0$, and

$$H = H_{appr} = E(0) > 0 \qquad (25)$$

in the limit $T \to 0$.

But according to Fig. 1 [1] $H_{appr} < 0$ at low temperatures. Therefore the data presented on Fig. 1 [1] for $H_{appr}$ could be incorrect, and the experimental data presented on Fig. 1 [1] for the enthalpy of the nitrogen molecule could be incorrect too.

5. The temperature dependence of the ratio $Q^v_{appr}/Q^v$ of the approximate vibrational partition function $Q^v_{appr}$ obtained from Eq. 11 to the exact one $Q^v$ obtained from Eq. 8 for the nitrogen molecule $N_2$ is presented on Fig. 1. One can see that the ratio increases from $1/2$ to $1$ with increasing temperature, and Eq. 11 gives incorrect values of the vibrational partition function, especially at low temperatures.

6. The temperature dependence of the relative difference $\Delta = (E^v_{appr}/E^v - 1) \cdot 100\%$ between the approximate mean vibrational energy $E^v_{appr}$ obtained from Eq. 13 to the exact one $E^v$ obtained from Eq. 12 for the nitrogen molecule $N_2$ is presented on Fig. 2. We established that $\Delta$ increases from $-50\%$ to $2\%$ with increasing temperature from zero to $3168.35\,K$, and it decreases from $2\%$ to $0\%$ with increasing temperature from $3168.35\,K$ to infinity, Eq. 13 underestimates the mean vibrational energy at temperatures lower than $1863.86\,K$, and Eq. 13 slightly overestimates the mean vibrational energy at temperatures higher than $1863.86\,K$.

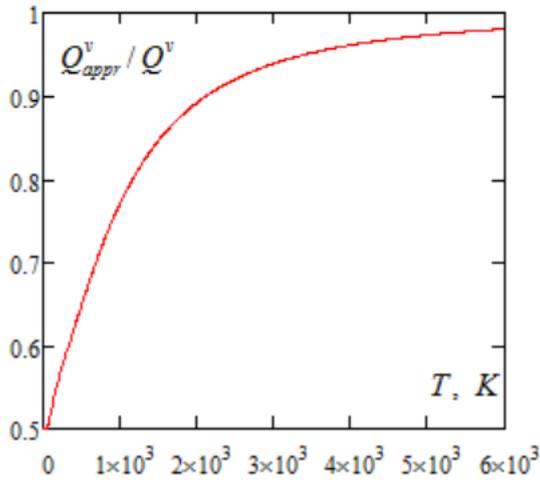

Fig. 1. Temperature dependence of ratio $Q^v_{appr}/Q^v$ of approximate vibrational partition function $Q^v_{appr}$ obtained from Eq. 11 to exact one $Q^v$ obtained from Eq. 8.

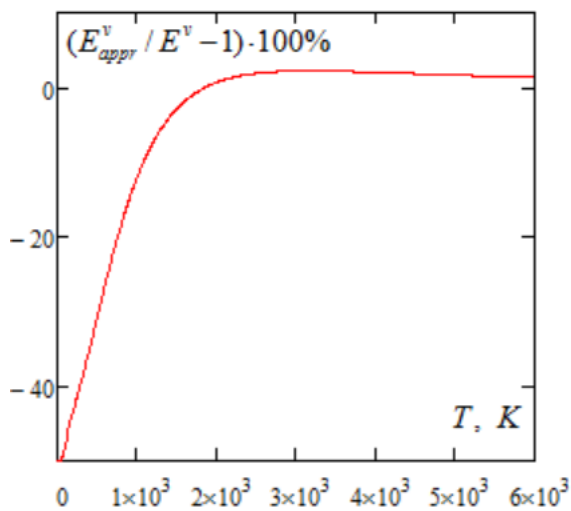

Fig. 2. Temperature dependence of relative difference $\Delta = (E^v_{appr}/E^v - 1) \cdot 100\%$ between approximate mean vibrational energy $E^v_{appr}$ obtained from Eq. 13 to exact one $E^v$ obtained from Eq. 12 for the nitrogen molecule $N_2$.

7. The temperature dependencies of the approximate $H_{appr}$ defined from Eqs. 13-16 and exact enthalpies $H$ defined from Eqs. 12 and 14-16 for the nitrogen molecule $N_2$ are presented on Fig. 3.

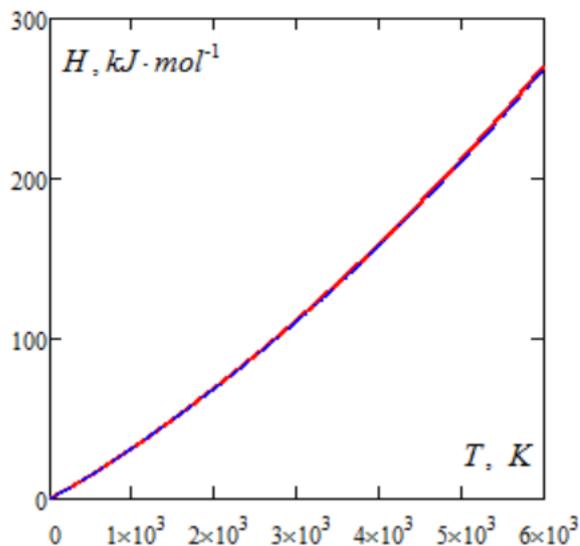

Fig. 3. Temperature dependencies of approximate $H_{appr}$ defined from Eqs. 13-16 (solid red line) and exact enthalpies $H$ defined from Eqs. 12 and 14-16 (dashed blue line) for the nitrogen molecule $N_2$.

As evident from Fig. 4 the values of the approximate and exact enthalpies for the nitrogen molecule are very close to each other despite the considerable difference between the approximate and exact vibrational enthalpies given by Eqs. 13 and 12, respectively, at temperatures below $1863.86\,K$.

As one can see from Fig. 4 the ratios of the approximate and exact vibrational enthalpies to the sum of the rotational and translational enthalpies are small. The smallness of these ratios at temperatures below $1863.86\,K$ is the reason of the good agreement between $H_{appr}$ and $H$.

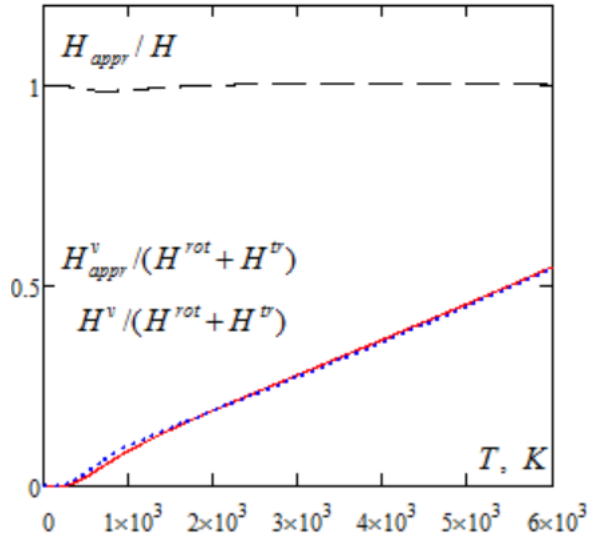

Fig. 4. Temperature dependencies of: ratio $H_{appr}/H$ of approximate enthalpy $H_{appr}$ defined from Eqs. 13-16 to exact enthalpy $H$ defined from Eqs. 12 and 14-16 for the nitrogen molecule $N_2$ (dashed black line); ratio $H^v_{appr}/(H^{rot}+H^{tr})$ of approximate vibrational enthalpy $H^v_{appr}$ defined from Eq. 13 to sum $H^{rot}+H^{tr}$ of $H^{tr}$ and $H^{rot}$ defined from Eqs. 15-16 (solid red line); and ratio $H^v/(H^{rot}+H^{tr})$ of exact vibrational enthalpy $H^v_{appr}$ defined from Eq. 13 to sum $H^{rot}+H^{tr}$ (dotted blue line).

**Conclusion**

It is shown that that the energy spectrum of the pure vibrational levels of the molecule consisting of two atoms interacting with each other via the modified Rosen-Morse potential, the analytical expressions for the vibrational partition function and enthalpy of the diatomic molecule obtained in [1] are incorrect.